# Balancing Personal Privacy and Public Safety during COVID-19: The Case of South Korea

**Na Young Ahn[1], Jun Eun Park[2], Dong Hoon Lee[3] and Paul C. Hong[4]**
[1]Institute of Cyber Security & Privacy at Korea University, Seoul, South Korea
[2]Department of Pediatrics, Department of Pediatrics, Korea University College of Medicine, Seoul, South Korea
[3]Institute of Cyber Security & Privacy & The Graduate School of Information Security at Korea University, Seoul, South Korea
[4]Information, Operations, and Technology Management College of Business and Innovation, The University of Toledo, USA

Corresponding authors: Dong Hoon Lee (e-mail: donghlee@ korea.ac.kr) and Paul C. Hong (e-mail: Paul.Hong@Utoledo.edu)

**ABSTRACT** There has been vigorous debate on how different countries responded to the COVID-19 pandemic. To secure public safety, South Korea actively used personal information at the risk of personal privacy whereas France encouraged voluntary cooperation at the risk of public safety. In this article, after a brief comparison of contextual differences with France, we focus on South Korea's approaches to epidemiological investigations. To evaluate the issues pertaining to personal privacy and public health, we examine the usage patterns of original data, de-identification data, and encrypted data. Our specific proposal discusses the COVID index, which considers collective infection, outbreak intensity, availability of medical infrastructure, and the death rate. Finally, we summarize the findings and lessons for future research and the policy implications.

## I. INTRODUCTION

Increasingly, the integration of big data and information and communications technology (ICT) promises enormous social value creation. In a pandemic crisis, public safety is a top priority. Simultaneously, we cannot ignore potential privacy breaches. The "old" debate over personal privacy and public security is still relevant. Since personal information is crucial to curtail the spread of a pandemic, policymakers and officials can more likely expect "implicit" consent. However, in the course of pursuing compelling public purpose, privacy rights may be at risk [1, 2, 3].

In general, any epidemiological study is subject to an ethics review to ensure privacy [4, 5]. For a face-to-face investigation, investigators respect confidentiality requirements. However, in view of the increasing social costs associated with the prevention and treatment of serious infectious diseases, there is a growing demand to gather accurate personal information in real time. For example, Gilbert Beebe suggested that in certain disastrous circumstances, public interest might be a higher priority than privacy issues [7]. The widespread flu epidemic in 2009 provided additional support for this line of reasoning. While conducting epidemiological investigations, researchers did not always obtain an individual's explicit consent. In the United States, the Health Insurance Portability and Accountability Act (HIPAA) established the privacy rules that set limits on the use and disclosure of any personal health information [8], though aggregating personal information for public health purposes is a somewhat different matter [9].

COVID-19 is an extraordinary circumstance that poses enormous public health risks—potentially affecting millions of people worldwide [10, 11, 12]. Nations are at war with the coronavirus. In this context, what does it mean to balance personal privacy and public safety? What are the boundaries and acceptable norms? This study considers these questions and examines the actual cases of two countries—South Korea and France. The subsequent sections of this study proceed as follows. In Section 2, we discuss the characteristics of the virus that causes COVID-19. We then introduce an anti-displacement alternative to COVID-19. We further compare the results of the French and Korean governments' quarantine measures against COVID-19. We apply the STRIDE Threat Model to perform a risk analysis of the Korean government's quarantine system. Our specific proposal considers collective infection, outbreak intensity, availability of medical infrastructure, and the death rate. Based on our findings, we present lessons and implications for future epidemiological investigations.

## II. COVID-19 RESPONSES

A new type of coronavirus (SARS-CoV-2) was first reported in Wuhan, China, in December 2019. Since then, this respiratory infection epidemic, designated as COVID-19, spread throughout China and worldwide. Upon infection, after a 2- to 14-day incubation period, patients experience respiratory symptoms including high fever (about 37.5 degrees) and cough or dyspnea [13-19]. However, it seems that there are several cases of asymptomatic infections. On January 21, 2020, the Chinese government officially reported 15 confirmed cases of COVID-19 [20-23]. The medical staff involved in the incident became credible evidence of human-to-human transmission [24]. On January 30, 2020, the World Health Organization (WHO) declared the continual spread of this infection as an International Public Health Emergency (PHEIC). With an accelerating rate of confirmed patients worldwide, on March 11, 2020 the WHO declared the COVID-19 outbreak a pandemic [25].

COVID-19 is a respiratory virus that spreads primarily through droplets generated when an infected person coughs or sneezes, or through droplets of saliva or discharge from the nose. The infected patient's saliva can be transmitted directly to another person's eye or if a person rubs their eyes with a virus-contaminated hand [24]. The rapid spread of COVID-19 was expected to overwhelm limited medical equipment and facilities with the sudden increase in the explosive number of patients [26]. Consequently, the fight against COVID-19 requires contact tracing for close contacts of laboratory-confirmed or probable patients. In some countries, these responses were compulsory, while others implemented a voluntary system. Our study compares the cases of France and South Korea, with special focus on the South Korean government's approaches in seeking the participation of its citizens.

### A. KOREAN GOVERNMENT'S APPROACH

At first, the South Korean government did not respond appropriately by not knowing the precise nature of the COVID-19 pandemic. The initial optimism was based on confidence that Korea's medical capabilities could handle any major public health challenges. Additionally, how to assess asymptomatic patients was determined somewhat later. For example, a Chinese woman who arrived from Wuhan on January 20, 2020 was identified as the first confirmed case. Until then, foreign tourists without fever were free to enter Korea, and there was not yet any serious effort to track asymptomatic patients.

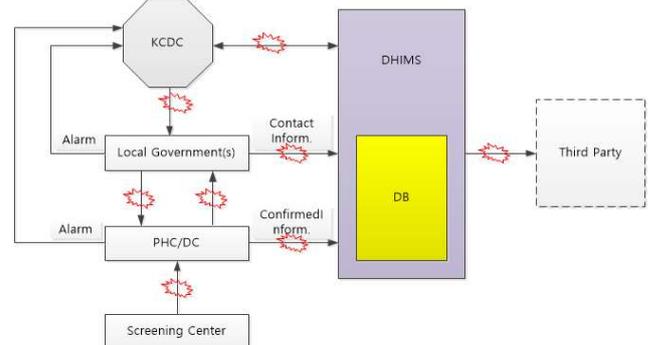

**FIGURE 1.** Disease Health Integrated Management System for COVID-19.

However, upon understanding the significance of asymptomatic patients and the nature of droplet infection, the next task was to identify the pathogens of confirmed patients. Fig. 1 describes the essential elements of the Disease Health Integration System (DHIMS), which collects and uses epidemiological survey data. Local governments conduct tests for epidemiologic investigation. Medical staff at public health centers and diagnostic screening centers follow-up with the confirmed patients. Local governments are responsible for operating screening clinics through large scale drive-through or walk-through testing sites without harvesting virus transmission [27].

If a person tests positive, then the health or diagnostic center immediately uploads the relevant personal information of the patient to the DHIMS. The health or diagnostic center immediately submits incident reports to the Korean Centers for Disease Control (KCDC). The local government health center also conducts an additional epidemiological investigation. The public safety law requires that confirmed patients disclose their recent movements and identify all contacted persons. The local government examines a confirmed patient's recent usage information from mobile phones and credit cards and uploads all information about the contacted persons including their name, address, contact information, date of birth, gender, disease name, diagnosis date, age, occupation, place of residence, telephone number, and health status, to the DHIMS. In this way, a national database of epidemiological investigations maintains all the relevant information of the confirmed patient and all contacts.

A diagnostic test is performed immediately for persons with symptoms. According to the severity of the COVID-19 symptoms, the individuals are either self-quarantines or hospitalized. Recent contacts who have no symptoms are quarantined for 14 days from the contact date of the confirmed patient. Self-quarantined individuals are monitored daily at local government call centers. If the additional diagnostic test after 14 days shows that the individual is negative and he/she has no symptoms, then the individual is released.

The Korean government implemented a COVID-19 response system with 3P (Preemptive, Prompt, and Precise) and 3T (Trace, Test, Treat) plus P (Participate) quarantine response model [27]. It used innovative ICT systems such as self-isolation and diagnostic apps, drive-through and walk-through clinics, and mobile phone location information analysis. The Korean government also counted on voluntary participation by citizens to develop additional app capabilities. For example, using these aggregated epidemiological survey databases, real time COVID-19 maps and monitoring apps were developed to benefit society at large.

### B. FRENCH GOVERNMENT'S APPROACH
France established Public Health France (PHF) on January 13, 2020 to monitor and respond to the COVID-19 epidemic [28, 29]. PHF's Crisis Center monitors epidemiological prevention, mobilizes health protection organizations, manages the strategic resources of medical facilities, and offers support services. The PHF conducts daily epidemiological investigations and releases the aggregate details including the area, gender, and age group of COVID-19 patients [30, 31].

The PHF set up a surveillance system to monitor the epidemiological and clinical aspects of COVID-19 using urban medicine, measure the severity of the epidemic and its impact on the medical system, and report the fatality rate. The PHF took active preventative measures to control the spread of COVID-19 with the aim of reducing the risk of transmission by providing warning messages to people in affected areas. In addition, the precautionary measures aimed to help people maintain a better quality of life, even in social isolation.

The PHF also supported active health-related services by operating a remote support system. The PHF allows healthcare professionals (e.g., doctors, nurses, pharmacists, physical therapists, midwives, etc.) and health professionals (managers, supervisors, health facility personnel, and engineers) to be prepared for a request for help from the health center. The French government implemented quarantine measures since March 16, 2020. The PHF monitors the behavioral responses and mental health practices in response to these changes and assesses social anxiety levels. Certainly, the COVID-19 pandemic disrupted the French ways of life and restricted vital economic and social activities. From the early days of the outbreak, the PHF's main challenge has been how to mobilize citizen participation in the fight against COVID-19.

### C. COMPARISONS OF FRANCE AND SOUTH KOREA
According to WHO data, the first confirmed case in South Korea was January 20, 2020. In France, the first reported case appeared on January 24, 2020. However, after little more than two months, these two countries showed a marked difference in terms of cumulative number of confirmed cases and total deaths [25]. As Table 1 (as of June 7, 2020) reports, the cumulative total of confirmed cases in Korea is 11,776 and 153,634 in France. The cumulative deaths are 273 and 29,143 for Korea and France, respectively.

TABLE 1
COMPARISON OF COVID-19 DATA FOR KOREA AND FRANCE

| Source Date: June 7, 2020 | France | South Korea |
|---|---|---|
| First Case | Jan. 24, 2020 | Jan. 20, 2020 |
| Lockdown | Mar. 16, 2020 | Feb. 18, 2020 |
| Total Confirmed Cases | 153,634 | 11,776 |
| Total Deaths | 29,142 | 273 |
| Population | 65,273,512 | 51,269,183 |
| Aged 65 Older(%) | 19.718 | 13.914 |
| Hospital Beds per Thousand | 5.98 | 12.27 |

Korea's population is 51,269,183, and France's population is 65,273,512 as of June 7, 2020 [25, 32]. In France, 20% of the population is over the age of 65, while it is about 14% in Korea, indicating that France has many elderly people. In addition, the number of beds per 1,000 people is 5.98 in France and 12.27 in Korea. In general, mortality is closely related to the number of hospital beds and the elderly population [33, 34]. Therefore, when comparing the hospital bed and proportion of the elderly population, more deaths should be more likely in Korea than in France. Both countries encouraged their citizens to join the fight against COVID-19. Even so, France has relatively more fatalities than Korea. These differences deserve careful analysis of other preventive measures. In the next section, we examine the impact of the epidemiological investigation database and the ICT technology usage in Korea.

## III. SECURITY OF THE KOREAN RESPONSE SYSTEM
It is beyond the scope of this study to describe the development of the Korean government's quarantine system processes and its operational mechanisms fully. For the purpose of this research, we applied available response guidelines released by the Korean government and explored other documents about the quarantine system [35, 36].

### A. THREAT ANALYSIS USING THE STRIDE THREAT MODEL
We evaluated security by applying the STRIDE threat model and examined the dynamic investigation data input and output of the DHIMS [27, 37].

Fig. 2 shows the sequence of on-line and off-line data collection, DHIMS storage, and third-party access. Potential data vulnerability spots are noted ( ) in the process linkage sequences. Threats to data integrity occur in several ways. Despite Korea's effective response to COVID-19 using epidemiological survey data, the entire process also contains potential privacy violations.

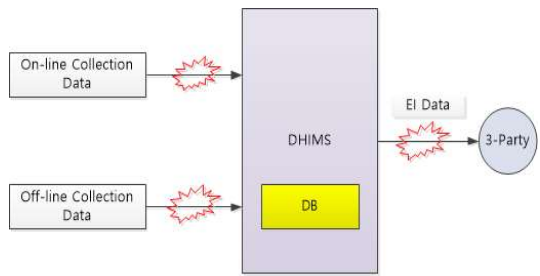

**FIGURE 2.** Potential Data Vulnerabilities in the System.

**Identity Spoofing:** The appropriate security level of the DHIMS system requires identity safeguarding and restricting access to epidemiological investigation data. After performing the basic authentication operation procedures, the relevant medical personnel, epidemiologists, government agencies, or civilians (i.e., third parties) are allowed to access the epidemiological investigation database. Identity spoofing; that is, misusing stolen identity data, is the most prevalent security risk attack.

**Data Tampering:** The on-line/off-line epidemiological data collection methods involve extensive personal information: name, address, contact information, gender, age, and phone number of the confirmed patient / all contacted. The DHIMS does not automatically de-identify such epidemiological survey data. During the data entry process, medical personnel or epidemiologists may make mistakes or arbitrarily change some of personal information content.

**Repudiation**: Because a legitimate user has access to the input and output of epidemiological survey data from the DHIMS, there must be sufficient correcting and checking procedures in the processing data operation. For example, what if the epidemiological investigation results in an offline state do not always exactly match the epidemiological data entered in an online state? Therefore, a repudiation option is necessary to ensure the integrity of the epidemiological investigation data.

**Information Disclosure**: The retention period for epidemiological data in DHIMS is permanent or semi-permanent. If the DHIMS has solid system security, then we can assume that all personal information is safe. However, when a third party requests a particular set of epidemiological investigation data, the DHIMS is supposed to conduct a de-identification process and offer specific numbers instead of names [37]. However, in the course of various information disclosures, an individual's privacy might not always be well-respected.

**Denial of Service/ Elevation of Privilege**: What might be problematic is the fact that these epidemiological investigation data have a legally permanent or semi-permanent retention period [37]. The DHIMS quality control measures require an application of relevant parameters (e.g., proper authorization and examination of usage patterns). Without the full operation of strict safeguarding measures when issuing permission or denial of personal information access, serous privacy risk concerns remain.

Table 2 summarizes the various types of threat and risk levels according to the DHIMS system access level.

TABLE 2
THREAT TYPE AND THREAT LEVEL

| Threat Type | Threat Level |
| --- | --- |
| Identity Spoofing | System Access Level |
| Data Tampering | Input of epidemiological investigation data: Medical personnel or epidemiologists can make any change |
| Repudiation | System Access Level |
| Information Disclosure | System Access Level<br>Epidemiological investigation data: privacy data, raw data<br>Providing epidemiological survey data to the government and the private sector |
| Denial of Service | System Access Level |
| Elevation of Privilege | System Access Level |

### B. PERSONAL PRIVACY VS. PUBLIC SAFETY

The Korean government disclosed the COVID-19 confirmatory movement paths and the addresses of quarantined buildings and enforced two weeks of self-containment for all confirmed patients and their contacts. In the early days of the epidemic, COVID-19 maps tracked the movements of these individuals, thus raising awareness of many people in affected areas.

In the digital age, balancing public safety and personal privacy is still enormously challenging [38, 39, 40]. With the rapid spread of COVID-19, unidentified aggregate information has little value. Public safety requires the "right to know" about the status of infection. Individuals may waive their privacy rights for public safety when it requires informing people about relevant COVID-19 infection information. The question is, "For legitimate public safety purposes, how do government authorities rightfully use personal information?"

### C. EXAMPLE OF BIG DATA USE IN THE COVID-19 PANDEMIC

From the early stage of the outbreak, the Korean government collected detailed personal information about the confirmed patients. Using these data (e.g., credit cards, phone number, and address), investigators could specify the paths of infection, conduct disaster prevention, and implement self-containment measures of all contacts. Such active follow-up methods had a considerable success.

On February 18, 2020, Korea had its 31st confirmed patient from the Shincheonji church in the Daegu area. With the sudden increase in confirmed patients among Shincheonji church members, the Korean government changed its approach and implemented more aggressive follow-up measures [27]. Shincheonji church, as a new religious movement, employ somewhat controversial elements in their recruitment of new members and the education of its existing members. In particular, their regular mass meeting often occurs in an enormous, enclosed hall. Hundreds of the church's leaders attended their international missionary outreach gathering in Wuhan, China and returned to Korea in January 2020. In the meantime, the number of confirmed patients increased explosively—up to 7,513 on March 10, 2020 [41, 42, 43].

Considering the rapid virus transmission among the church members, the Korean government took aggressive action. At the government's request, the Shincheonji Church provided the social security and phone numbers of its members. Local governments called the church members in their region, looked for symptoms, and conducted COVID-19 tests. The Shincheonji church ledger has more than 200,000 people. Thus, nearly all the church's members (about 212,000) were contacted and examined. The Korean government used this set of big data to prevent the COVID-19 pandemic. With the use of this data and follow-up testing, the government effectively contained one of the main sources of the widespread outbreak.

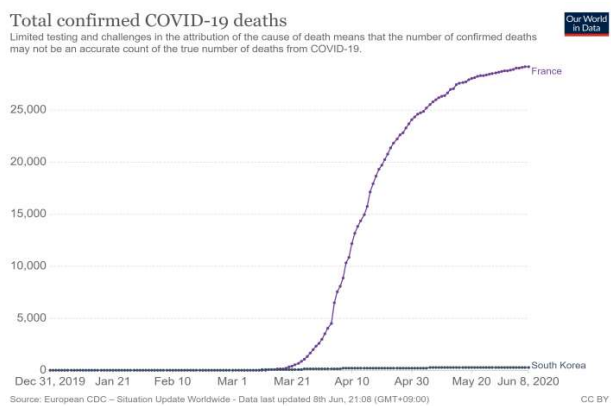

**FIGURE 3.** Comparison of cumulative deaths in South Korea and France [44]

Fig. 3 shows a comparison of the cumulative deaths in Korea and France. The number of deaths before and after the COVID-19 pandemic, declared on March 11, shows a sharp difference. As we discussed in the case of Shincheonji church, Korea's aggressive and extensive use of personal information made a significant difference. The question is, "What is the proper way to use personal information, even in this pandemic?" Here, we consider the value of using de-identified information.

### D. DE-IDENTIFICATION IN EPIDEMIOLOGICAL INVESTIGATIONS

As a proactive measure to contain and prevent the pandemic, the Korean government used epidemiological data through ICT. Medical testing equipment expedited massive COVID-19 testing, which was instrumental in reducing mortality. It is a public safety imperative to use personal information to control and prevent the spread of a pandemic. Managing personal information stored in big data sets requires appropriate privacy protection measures. Privacy violation is related to the use of identifiable personal information. Therefore, an effective safeguard of personal privacy requires the use of non-identifiable personal information. The right type of technological support is essential for such de-identification options.

In the United States, the HIPPA set national standards for the protection of an individual's medical records and personal health information. It applies to health plans, health care information centers, and health care providers that transmit any health care transactions electronically. This rule requires appropriate safeguards to protect the privacy of personal health information, sets limits, and specifies the conditions for the use and disclosure of such information without patient consent and approval [9, 45].

However, the Korean government uses identifiable personal information with limited restrictions, potentially leading to serious violations of the privacy of patients and their contacts. Securing personal information for quarantine measures is appropriate. With respect to personal privacy, it is important to use the information gathered for the specific intended purpose only. Using identifiable personal information for any other purpose is a breach of confidence and trust. Moreover, the legal provision of keeping quarantine investigation data either permanently or semi-permanently is not reasonable at all. Requiring the de-identification of personal information and rapid-deployment-relevant technology is an urgent need [46].

## IV. ADAPTIVE EPIDEMIOLOGICAL INVESTIGATIONS

The purpose of conducting epidemiological investigations is to understand the nature of an epidemic and determine how to control the spread of infectious diseases for public safety [47]. However, public safety does not justify privacy infringement. In this section, we discuss the practical steps epidemiological investigations can take to achieve a balance between privacy and public health.

### A. CLASSICAL TRADE-OFF BETWEEN PERSONAL PRIVACY AND PUBLIC SAFETY

The HIPAA's privacy rules propose two approaches to the de-identification of personal health information: the safe harbor method and the expert determination method [45]. The safe harbor method deletes 18 personal identification variables such as name, social security number, contact information, IP address, fingerprints, photographs, and detailed address. The method of using experts is to process personal information using non-identifying algorithms.

The release and forget model, the data use agreement (DUA) model, and the enclave model are all useful to achieve effective control in data storage and usage processes. The general pre-sale model is to release unidentified personal information to the public by posting data online. A DUA establishes sharing rules between research collaborators who are covered entities under the HIPAA privacy rule. Only intended recipients can use certain information in a limited data set. The closed room model maintains a safe analytic environment that restricts unauthorized access and the export of personal information in its original form. It is a physical and technical control method to respond and export [48].

It is often challenging to enhance the scientific utilization value of data collected with de-identified personal information. An increasing level of de-identification is negatively related to the quality of the data and the precision of the research results. Conversely, higher data quality and outcome precision require lower levels of de-identification. A greater level of personal identification is related to a higher possibility of privacy infringement [49].

Therefore, individual researchers aiming to achieve more precise analysis results may prefer to use the original data, which contains identifiable personal information. On the other hand, reputable institutions satisfy personal privacy requirements by ensuring the anonymity of the data.

### B. BALANCING PERSONAL PRIVACY AND PUBLIC SAFETY

There are diverse approaches to data de-identification. In general, methods to determine the level of de-identification of personal information are known [48]. The National Institute of Standards and Technology (NIST) proposes a method determined by an expert and a safe hopper method for removing multiple identifiers [50, 51]. In recent years, differential privacy technology that adds noise to personal information has been attracting attention as a way to increase privacy in big data analysis [52, 53]. This differential privacy technique is a de-identification method that performs a kind of pseudonymization process.

With widely available de-identification technologies, it is difficult to prevent individuals from being re-identified from de-identification measures. Researchers at the Imperial College London, UK, conducted experiments with published data from the United States, Turkey, and other countries, and found certain attributes accurately even by using de-identified data [54]. Their machine learning model could identify individuals with 99.98% accuracy from anonymized data using only 15 demographic attributes (age, gender, marital status, etc.). This research team suggests a paradigm shift: "We need to de-identify, then move on. Anonymity is not a property of the data set. It depends on how the person who writes it uses it." In other words, what matters is not anonymizing, but designing and organizing data to be useful and meaningful.

So then, what are the alternatives? It is time to move from the idea of de-identification to the application of appropriate technologies. The aim is to strike a balance between the public use of data and personal privacy [55, 56]. Technologies such as secure multi-party computation and homomorphic encryption are emerging. More innovations are certainly in progress in the post-COVID-19 world of new big data technologies and ICT applications [57, 58].

The development of these new technologies can increase our options when dealing with infectious diseases. It is imperative to balance personal privacy and public safety, even in the context of COVID-19. Personal information with individual consent may be used for specific research purposes.

### C. THE NEED FOR AN INDEX TO BALANCE PERSONAL PRIVACY AND PUBLIC SAFETY

There has been serious debate over the value priorities of the epidemiological investigation. Healthcare policymakers are more likely to lean toward public safety goals. On the other hand, safeguarding personal privacy is important from the individual rights perspective. In this context, developing an effective mechanism for balancing public safety and personal privacy is important and timely. We present an index measure for balancing criteria. Our study provides a helpful practical tool in epidemiological investigations.

### D. COVID INDEX FOR EPIDEMIOLOGICAL INVESTIGATION

By applying the concept of DREAD modeling in security engineering, we propose the COVID index as a method to balance public health and privacy in epidemiological investigations [59]. This COVID model uses five parameters: collective infection, outbreak intensity, viral tiler, infrastructure of medical faculties (e.g. number of medical beds per million people), and death rates (or fatality rate).

TABLE 3
COVID INDEX FOR INTELLIGENT EPIDEMIOLOGICAL INVESTIGATIONS

| Parameter | Low | High |
|---|---|---|
| C (Collective infection) | 0 | 5 |
| O (Outbreak intensity) | 0 | 5 |
| V (Viral tiler) | 0 | 5 |
| I (Infrastructure of medical faculties) | 0 | 5 |
| D (Death rates) | 0 | 5 |

Table 3 illustrates how adaptive epidemiological investigations may use the COVID index. Here, C represents collective infection, O represents the outbreak intensity, and V represents the viral propagation power. Here, the value indicates the minimum concentration at which the virus infects the cell. I represents the level of medical infrastructure. D represents the mortality rate of the virus. The COVID index can be calculated as follows:

$$\text{COVID index} = \frac{C+O+V+I+D}{5}. \quad (1)$$

Values from 0 (low) to 5 (high) are assigned to each item. The values of each item are summed according to Equation 1, and the COVID index is determined as the average value of the results. The sum with be a value from 0 to 25, and the COVID index will therefore be a value 0, 1, 2, 3, 4, or 5. A high COVID index suggests a significant risk of the virus's propagation power and public health and an urgent need to investigate the epidemiology. In this case, aggressive epidemiological investigations should be conducted by collecting original data. Aggressive epidemiological investigations minimize the incidence of additional confirmed patients from contact and suspected patients [60, 61]. Quarantine measures can rapidly contain a virus. Thus, deploying the available medical resources has the maximum prevention effect.

If the COVID index is greater than or equal to 2, then the epidemiological investigation should focus on collecting and using de-identified data. On the other hand, for a COVID index of either 0 or 1, researchers should collect and use encrypted data instead.

### E. SUGGESTIONS TO STRENGTHEN PRIVACY IN EPIDEMIOLOGICAL INVESTIGATIONS

The primary purpose of the epidemiological investigation is to minimize contact with a confirmed patient. Thus, isolating individuals that test positive for a disease is imperative to prevent the occurrence and spread of infectious diseases. Here, we suggest several practical suggestions to enhance security in the epidemiological investigations.

First, investigators should be required to obtain a personal consent forms and use personal information within a specific period. In the early breakout period of COVID-19, personal information was often collected without a proper personal consent process. Later, a mandatory requirement to specify the data storage period and usage patterns of personal information was put in place. If proper consent forms are not obtained, then the personal information collection process should stop. All personal information stored in the database should include the entry time and expiration date. The investigation system should automatically delete epidemiological survey data after the expiration date. Further steps may also be taken to remove personal information completely from other databases and thus guarantee personal privacy [62].

Second, we should explore other options to use identifiable personal information. What if de-identification is not practical for research purposes? Personal information is regarded as similar to the copyright concept. For the sale of any products with copyright, a certain amount of money is set aside to compensate the copyright holder. Similarly, it is plausible to compensate each individual for the use of their personal information for specific research purposes.

Third, we should address de-identification technology. An individual from the medical field or an epidemiologist may apply de-identification technology when storing personal information. When the required information is collected through off-line systems and then uploaded to a database, then the individual under investigation should be notified to check the accuracy and provide consent. Afterward, the offline information should be destroyed immediately and the individual should be notified of the destruction. Epidemiologists should do the same when they apply de-identification technology to store personal information collected online.

Fourth, researchers should establish conditions for third party access. Any personal information provided to a third party should be made available in the form of non-identifying numbers or symbols. In case a third party need to use identified personal information, they should require personal consent.

Fifth, researchers should design an operating system for personal privacy. Google and Apple recently released a tracking system with privacy features [63, 64]. Other scholars also introduced systems that encrypt data to ensure privacy in applications [65, 66]. These options offer additional safeguards for ensuring personal privacy.

Adaptive epidemiological surveys may still contain human errors in the course of using different types of technologies, including artificial intelligence (AI)-based epidemiological investigation systems. Implementing the suggestions above should aim to improve personal privacy in epidemiological investigations. In addition, our proposed COVID index can provide a basis for epidemiological investigations to support efforts to balance personal privacy and public safety.

### F. IMPROVING THE INTEGRITY OF OFFLINE DATA IN EPIDEMIOLOGICAL INVESTIGATIONS

People issues are too often related to data integrity and information quality. In epidemiological investigations, offline information gathering raises questions about the reliability. Incorrect information obtained from interviews with patients may lead to wrong assessment and evaluation about quarantine decisions. Therefore, it is important to check the quality and assure the integrity of offline epidemiological investigations. Specific security measures we propose are to strengthen an epidemiological investigation system. It is to cross-check the accuracy of offline information in real time using other online information sources (e.g., usage history of credit cards, bus and subway transportation cards). This will enhance the integrity of data gathering process. It will also prevent the rapid spread of infectious diseases through monitoring of the history of patients contacts and taking additional preventive measures for all those affected.

### V. CONCLUSION

In the COVID-19 context, the Korean government actively used personal information and achieved fairly successful public safety outcomes. However, that is only part of the whole story. The extensive use of personal information may also negatively impact personal privacy. Therefore, practical safeguard measures, including clear communication of the scope of public disclosure and the de-identification of personal information are required. This paper examined how to implement personal consent procedures and the appropriate use of big data. Even in a devastating pandemic like COVID-19, balancing personal privacy and public safety is still very important. Future research may explore how to prepare for other pandemic outbreaks by combining the capabilities of governmental leadership, technological innovation, big data use, and societal cooperation. However, such aggressive epidemic control measures involve personal privacy concerns. Further investigations should consider cultural issues related to privacy and public safety in different national contexts.

### NOTE of APPRECIATION

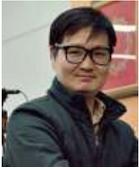

**Na Young Ahn** is a post-doc researcher with the Institute of Cyber Security & Privacy at Korea University, South Korea. He holds a Ph.D. in Cyber Security. He received his B.S. and M.S. degrees from the Department of Electrical Engineering at Korea University. He has been a patent engineer at patent and law firms since 2005. His articles have been published in journals including IEEE Access and Ad hoc & Sensor Wireless Networks. His research interests include physical layer security in vehicular communications, biometric authentication, PoN based blockchain, and anti-forensics in flash memories.

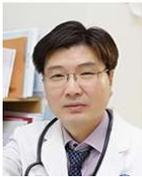

**Jun Eun Park** is a medical doctor (M.D.) majoring in pediatric hematology-oncology and hematopoietic stem cell transplantation. He graduated from Korea University College of Medicine in 1991 and completed an internship, and from 1994 to early 1999, he completed resident course at Asan Medical Center Hospital in Seoul. After working at Dankook University Hospital until 2003, he served as a professor of pediatrics at Ajou University School of Medicine from 2003 to mid-2020, and is now a professor of pediatrics at Korea University College of Medicine. He served as the educational director of the Korean Society of Pediatric Hematologic Oncology (KSPHO) from mid-2020 to September 2015 to August 2017, and served as the Wilms Oncology Subcommittee in the Korean Pediatric Hematologic Oncology Group (KPHOG). From 2017 to 2019. From July 2019 to the present, he is the chairman of the Korean Society of Pediatric Neuro-Oncology (KSPNO).

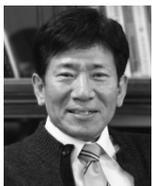

**Dong Hoon Lee** received his B.S. degree in economics from Korea University, Seoul, Korea, in 1985 and M.S. and Ph.D. degrees in computer science from the University of Oklahoma, Norman, OK, USA, in 1988 and 1992, respectively. Since 1993, he has been with the Faculty of Computer Science and Information Security, Korea University. His research interests include the design and analysis of cryptographic protocols in key agreement, encryption, signatures, embedded device security, and privacy-enhancing technology.

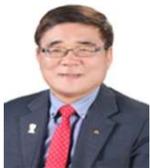

**Paul C. Hong** is a Distinguished University Professor of Global Supply Chain Management and Asian Studies at the University of Toledo, USA. He received his B.A. in Economics from Yonsei University, South Korea and an M.A. in Economics and MBA from Bowling Green State University, USA, and a Ph.D. in Manufacturing Management and Engineering from the University of Toledo, USA. His articles have been published extensively in journals including the Journal of Operations Management, Journal of Supply Chain Management, International Journal of Production Research, International Journal of Production Economics, Journal of Engineering Technology Management, British Journal of Educational Technology, and European Management Journal. His recent books include Rising Asia and American Hegemony (2020; Springer) and Creative Innovative Firms (2019; Springer). His research interests are in global supply chain management, entrepreneurial innovation, and the interfaces of ToP and BoP.